\documentstyle[prd,preprint,aps]{revtex}
\begin{document}

\reversemarginpar
\tighten

\title{Entropy bounds for charged and rotating systems} 
\author{Gilad Gour\thanks{E-mail:~gilgour@phys.ualberta.ca}}
\address{Theoretical Physics Institute, 
Department of Physics, University of Alberta,\\
Edmonton, Canada T6G 2J1}

\maketitle

\begin{abstract} 
It was shown in a previous work that, for systems in which the entropy
is an extensive function of the energy and volume, the Bekenstein and 
the holographic entropy bounds predict new results. In this paper, we
go further and derive improved upper bounds to the entropy of
{\it extensive} charged and rotating systems. Furthermore, 
it is shown that for charged and rotating systems (including
non-extensive ones), the total energy that appear in both the
Bekenstein entropy bound (BEB) and the causal entropy bound (CEB) can
be replaced by the {\it internal} energy of the system. In addition, 
we propose possible corrections to the BEB and the CEB.
\end{abstract}

\pacs{PACS numbers:~}

\section{Introduction}
According to classical general relativity, the second law of
thermodynamics is violated when a system crosses the event horizon of
a black hole. This had led Bekenstein~\cite{Bek73} to conjecture that
the area of a black hole (in suitable units) may be regarded as the
black hole entropy. In addition, Bekenstein proposed to replace the
ordinary second law of thermodynamics by the 
{\it Generalized Second Law} (GSL):  
The generalized entropy, $S_{g}\equiv S+S_{BH}$, of a system consisting 
of a black hole (with entropy $S_{BH}$) and ordinary matter (with
entropy $S$) never decreases with time (for an excellent review on the
thermodynamics of black holes and the validity of the GSL, see
Wald~\cite{W01}). 

It is not very clear, however, if the validity of the GSL depends on
the existence of a universal upper bound to the entropy of a bounded
system: 
Consider a {\it gedanken-experiment} such that one lowers adiabatically 
a spherical box of radius $R$ toward a black hole (Geroch process). 
The box is lowered from infinity
where the total energy of the box plus matter contents is $E$.
It was shown~\cite{Bek81} that the entropy $S$ of the box must obey
(throughout the paper $k_{B}=1$)
\begin{equation}
S\le {2\pi R E\over\hbar c},
\label{bb}
\end{equation} 
in order to preserve the GSL. Recently, universal entropy bounds for
charged and rotating systems have been proposed by several 
authors~\cite{Hod1,BekMayo,Hod2,Mayo,Linet}.

However, the naive derivation of Eq.~(\ref{bb})  
in~\cite{Bek81} was criticized by Unruh and Wald~\cite{UW82,UW83} who
have argued that, since the process of lowering the box is a quasi-static
one (and therefore can be considered as a sequence of static-accelerating 
boxes), the box should experience a buoyant force due to the Unruh 
radiation~\cite{Unr76}.
Describing the acceleration radiation as a fluid, they have shown that this 
buoyant force alters the work done by
the box such that no entropy bound in the form of Eq.~(\ref{bb})
is necessary for the validity of the GSL. A few years ago, 
Pelath and Wald~\cite{PW} gave further arguments in favor    
of this result. 

Bekenstein~\cite{Bek83,Bek94}, on the other hand, argued that, only 
for very flat systems, the Unruh-Wald effect may be important. Later on, 
he has shown~\cite{Bek99} that, if the box is not almost at the horizon, the 
typical wavelengths in the radiation are larger than the size of the box
and, as a result, the derivation of the buoyant force from a fluid picture 
is incorrect.    
The question of whether the Bekenstein bound follows from the GSL via
the {\it Geroch process} remains controversial 
(see~\cite{And99,W01,Bek01,MS02}). 
However, as it was shown by Bousso~\cite{BouBek} (see the following paragraphs), 
there is another {\it link} connecting the GSL with the Bekenstein bound.

Susskind~\cite{Sus95} has shown, by considering the conversion of a 
system to a black hole, that the GSL implies a spherical entropy bound
(SEB)
\begin{equation}
S\le {1\over 4l_{p}^{2}}A,
\label{sb}
\end{equation}
where $S$ is the entropy of a system that can be enclosed by a sphere
with area $A$. A few years later, Bousso~\cite{Bou99,Bou02} had found 
an elegant way to generalize Eq.~(\ref{sb}) and write it in a covariant
form. He proposed the {\it covariant
entropy bound}: ``the entropy on any light-sheet $L(B)$ of a surface $B$
will not exceed the area of $B$". That is,
\begin{equation}
S[L(B)]\le\frac{A(B)}{4l_{p}^{2}},
\label{ceb}
\end{equation}
where the light-sheet $L[B]$ is constructed by the light rays that emanate
from the surface $B$ and are not expanding (for an excellent review
see Bousso~\cite{Bou02}). 

Recently, Flanagan, Marolf and Wald~\cite{FMW} have generalized
Eq.~(\ref{ceb}) into the following form:
\begin{equation}
S[L(B,B')]\le\frac{A(B)-A(B')}{4l_{p}^{2}},
\label{gceb}
\end{equation} 
where $L(B,B')$ is a light-sheet which starts at the cross-section
$B$ and cuts off at the cross-section $B'$ before it reaches a
caustic.
They where motivated by the argument that when a matter system with
initial entropy $S$ falls into a black hole, the horizon surface area
increases at least by $4l_{p}^{2} S$ due to the GSL. 

Unlike the controversial issues regarding the relationship between the 
GSL and Eq.~(\ref{bb}), the entropy bounds in 
Eqs.~(\ref{sb},\ref{ceb},\ref{gceb}) are closely related to the GSL.
However, very recently, Bousso~\cite{BouBek} has shown that 
the BEB follows from Eq.~(\ref{gceb}) for any
isolated, weakly gravitating system. Hence,  
even though it is not clear whether quantum effects should be taken
into consideration in the derivation of Eq.~(\ref{bb}) (via the Geroch 
process), there is a strong link between the GSL and the Bekenstein 
bound. 

In a previous work~\cite{Gour1}, it was shown that there is another
link connecting the bound~(\ref{bb}) with the entropy of thermal
radiation and the Stephan-Boltzmann law. In our derivation, we have
considered systems in which the entropy is an extensive function of
the energy. We have also showed that for such systems, the 
SEB~(\ref{sb}) yields the {\it causal entropy bound} (CEB)
proposed by Brustein and Veneziano~\cite{BV00} and by
Sasakura~\cite{Sasa}. In the present paper, 
we generalize our results for charged rotating systems. The importance
for such generalizations stems from the following.
 
The entropies of closed systems in flat spacetime are usually much
smaller than the entropy bound~(\ref{bb}) or its generalizations for
charged and rotating systems~\cite{Hod1,BekMayo,Hod2,Mayo,Linet} 
(see also section III and IV). 
Furthermore, since these bounds are not extensive functions of
the energy, one can expect that there are much tighter bounds that
are applicable only for extensive systems. As we shall see in this paper
(see also our previous work~\cite{Gour1}), such extensive bounds follow
from the BEB and also from the SEB. Thus,
in order to test the Bekenstein, as well as the spherical, entropy
bounds, it is enough to test whether their predictions on extensive
systems are valid.

For example, in our previous work, it was shown that the
BEB~(\ref{bb}) implies that the entropy of extensive spherical
systems can not exceed $(ER/\hbar c)^{3/4}$ (up to numerical factor).
Since  the entropy of thermal radiation
saturates this extensive bound, it is a good guess that no other
system has more entropy~\footnote{The rest mass of 
ordinary particles only enhances gravitational instability without
contributing to the entropy}. However, if one can find an extensive
system which exceeds this bound, it will be a counter example to the
BEB. As we shall see here, the extensive entropy
bound of charged rotating systems is tighter than $(ER/\hbar c)^{3/4}$
and therefore it provides a new challenge on the non-extensive entropy
bounds.

This paper is organized as follows: In section II we discuss the
validity of the BEB and we propose to include a
logarithmic correction term. In sections III and IV, we derive 
new entropy bounds for extensive charged and rotating systems
(i.e. systems with relatively small charge and small angular
momentum). This bounds turn out to be much smaller than the
non-extensive ones. In section V, the 
SEB is applied to extensive charged-rotating
systems and new extensive bounds are
proposed. In section VI we summarize our results and consider
non-extensive systems. We propose improved entropy bounds for charged
and rotating systems that generalize both the BEB and the CEB in a
natural way. We also obtain a correction term to the CEB.    

\section{The validity of the BEB}

Beside the question whether the bound~(\ref{bb}) is needed for the the
validity of the GSL, one may ask under what conditions it does hold. For
composites of non-relativistic particles the bound is trivially satisfied
since the entropy is of the same order as the number of particles
involved. As we shall see now, there is one exception to this
argument, but also then the bound is satisfied due to some other
physical arguments. First, we would like to confirm that a system of
$N$ non-relativistic particles, each with mass $m$, cannot have entropy
greater than the BEB. 

Let us denote the {\it kinetic} energy of the system by
$E_{k}$. Since the system is composed of non-relativistic particles we
assume that the entropy of the system is a function of
$\hbar$, $m$, $E_{k}$, $R$ and $N$. Since the entropy can be written
as a function of dimensionless quantities, we infer that it
is a function of {\it only} two independent parameters: $N$ and
$w\equiv m E R^{2}/\hbar ^{2}$. Next, we assume that the chemical
potential of the system vanishes. In that case the entropy is a
function of $w$ only, and since it is also an extensive function of
$E$ and $V={4\pi\over3}R^{3}$ it must be proportional to
$w^{3/5}$. That is, the entropy is proportional to
$E_{k}^{3/5}R^{6/5}$. 

The kinetic energy, $E_{k}$, is smaller than the total energy, $E$, that appears
in the BEB; approximately, $E_{k}=E-Nm$. However, at
first sight, it looks that by taking $R$ to be large enough such that
$R^{6/5}\gg R$ (ignoring the dimensions), one would be able to exceed the BEB. 

The system described above might represent, for example, a degenerate Bose gas. 
Below the critical temperature $T_{0}\sim (N/V)^{2/3}$ the entropy is indeed
proportional to $E_{k}^{3/5}R^{6/5}$. However, this is true only 
{\it below} the critical temperature. Thus, for $T_{0}>T$, $R$ is
bounded from above\footnote{Here we assume that $T$ remains constant. Otherwise, 
by increasing $R$ and decreasing $T$ (such that $T<T_{0}$), the kinetic energy 
is decreased as well. Thus, it can be shown that the decrease in $E_{k}$ is enough 
to protect the BEB.} and the entropy of the system can not exceed the
BEB. This example illustrates that without the
considerations of the actual physical system that exist in nature,
mathematically, it is very easy to find counter examples for the
BEB.       

For free
massless quantum fields enclosed in volumes of various shapes the
bound's validity has been checked directly (see review by Bekenstein
and Schiffer~\cite{BekSch}). 
Nevertheless, in~\cite{Gour1}, it was shown that the
bound impose a restriction on the number, $n_{s}$ of massless fields
to be no more then $\sim 10^{4}$.\footnote{The limitation
on the number of massless species in nature is a consequence of any
sort of entropy bound.} 
 
In the derivation of the bound by Bousso~\cite{BouBek}, three
assumptions have been made: weak gravity, the system is enclosed in a
spatially compact 
region and the null energy condition. It is the second assumption that
we would like to emphasize here. 
This assumption implies that $E$ includes the entire system. As shown
by Page~\cite{Page}, if this requirement is not satisfied one can find
counter examples to the BEB. 

According to quantum mechanics, the Compton wave length of the system
is given by $\lambda =\hbar c/E$. Thus, the system will satisfy Bousso's
second assumption if $ER\gg\hbar c$. Let us define the validity domain
of the BEB as follows:
\begin{equation}
\frac{ER}{\hbar c}\geq\gamma,
\label{defr}
\end{equation}
where $\gamma$ is a dimensionless constant of order unity. In order to
determine $\gamma$ one has to know how much energy is allowed to leak
out of the box. Furthermore, $\gamma$ determines the minimum possible
value of the BEB. That is, the BEB implies that $S\leq 2\pi\gamma$ for 
systems with radius $R=\gamma\hbar c/E$. Hence, a system that saturates 
the bound at this limit will provide information on $\gamma$ (and vice
versa).    
      
In curved (non-spherical) spacetime it is not very
clear how to define ``E'' and ``R''. However, for {\it spherical}
self-gravitating systems, Sorkin, Wald and Jiu~\cite{SWJ} provided an
indication 
that probably some version of the BEB may hold. If that
is correct, one would expect logarithmic corrections to the BEB.

The leading order corrections to the Bekenstein-Hawking entropy are
presumably logarithmic (see~\cite{Gour2} and references
therein). That is,
\begin{equation}
S_{BH}=\frac{A}{4l_{p}^{2}}-k\log\left({A\over l_{p}^{2}}\right)+...
\label{BH}
\end{equation}
where there are few indications that $k=3/2$. Therefore, the
SEB may have the same corrections. As a result,
the BEB~(\ref{bb}) will exceed
$A/4l_{p}^{2}-k\log(A/4l_{p}^{2})$ in the limit when 
$E \rightarrow c^{2}R/2G$. This motivates us to introduce a logarithmic
correction to the BEB:
\begin{equation}
S\leq\frac{2\pi ER}{\hbar c}-k\log\left({ER\over \hbar c}\right).
\label{modbek}
\end{equation}
Since $ER/\hbar c$ is usually much smaller than $A/l_{p}^{2}$, the
corrections to the Bekenstein bound are relatively more important
than the corrections to the entropy of a black hole. 

\section{An extensive entropy bound for charged systems}

About ten years ago Zaslavskii~\cite{Zas} has suggested how to tighten
Bekenstein's bound on entropy when the object is electrically
charged. Recently, Bekenstein and Mayo~\cite{BekMayo} have derived the
bound by considering the accretion of an ordinary charged object by 
a black hole. They have found that the upper bound to the entropy $S$ 
of an arbitrary system of proper energy $E$, proper radius $R$, and
charge $Q$:  
\begin{equation}
S\leq\frac{2\pi ER}{\hbar c}-\frac{\pi Q^{2}}{\hbar c}.
\label{ebc}
\end{equation}

The entropy bound~(\ref{ebc}) is understandable from a dimensional
point of view. That is, an entropy $S$ is a dimensionless function. 
For (charged) systems in flat space time, $S$ does not depend (explicitly) 
on the gravitational constant $G$. Furthermore, if the entropy bound
does not depend on the mass of some particular species, it can be
written in the form $S=f(x,y)$ where $x\equiv 2\pi ER/\hbar c$,  
$y\equiv \pi Q^{2}/\hbar c$, and $f$ is some arbitrary function of $x$
and $y$ (any other dimensionless quantity of $E,R,Q,\hbar,c$ must be a
function of $x$ and $y$). Now, the spherical entropy bound implies
that
\begin{equation}
f(x,y)\leq \frac{\pi}{l_{p}^{2}}R^{2}\equiv z,
\label{20}
\end{equation}
where $x,\;y$ and $z$ are all {\it independent} dimensionless
parameters. Therefore, one can fix $x$ and $y$ and take $z$ to its
minimal value. The minimal value of $z=z_{min}(x,y)$ is obtained when
the system becomes a charged black hole with definite $x$ and
$y$. Therefore, the expression for the radius of a Reissner-Nordstrom 
black hole implies that $z\geq z_{min}=x-y$. By substituting the minimum value 
of $z$ in Eq.~(\ref{20}), one obtains the bound~(\ref{ebc}).

Next, we would like to consider applications of this bound to extensive
systems (i.e. systems in which the energy and charge are
distributed uniformly). Since the energy for highly charged systems
can not be distributed uniformly, we assume that the Coulomb energy, $E_{c}$,
is much smaller than the total energy, $E$.  
When the charge is uniformly spread on
a ball of radius $R$, $E_{c}=3Q^{2}/5R$. Thus, in this section $E\gg
Q^{2}/R$. 

The entropy is a function of the {\it internal} energy
$E_{in}=E-3Q^{2}/5R$. Therefore, the bound~(\ref{ebc}) implies
\begin{equation}
S(E_{in})\leq \frac{2\pi R E_{in}}{\hbar c}+\frac{\pi Q^{2}}{5\hbar c }
\end{equation}
where we have substituted $E=E_{in}+3Q^{2}/5R$. However, since the
entropy does not depend explicitly on $Q$, we infer that
\begin{equation}
S(E_{in})\leq \frac{2\pi R E_{in}}{\hbar c}.
\label{nonextb}
\end{equation} 

Now, for systems in which the entropy is an extensive function of
$E_{in}$, one can write
\begin{equation}
S(E_{in},...)=V s(\varepsilon _{in},...),
\label{extcon}
\end{equation}
where $\varepsilon _{in}\equiv E_{in}/V$ is the 
internal energy density and $s$ is the entropy density (the dots
in Eq.~(\ref{extcon}) indicate that the entropy might depend on other
extensive parameters such as the number of particles etc). The
entropy bound in the right hand side of Eq.~(\ref{ebc}) does not
satisfy the above condition. Hence, we are motivated to seek  
a tighter bound for extensive systems (i.e. systems in which
the entropy is an extensive function of energy, charge etc). Indeed,
as we shall see now, such a bound exist and has some interesting
features.

Applying the entropy bound~(\ref{nonextb}) on extensive systems leads to the
following bound on the entropy density of the system:
\begin{equation}
s(\varepsilon _{in},...)\leq\frac{2\pi\varepsilon _{in}}{\hbar c}R.
\label{extb}
\end{equation}
Since $\varepsilon _{in}$ and $R$ can be considered to be
independent, one can fix $\varepsilon _{in}$ and change
$R$. However, there are limitations on the range allowed to $R$. First,
as we have mentioned earlier, the Coulomb energy $Q^{2}/2R$ is much
smaller then the {\it total} energy $E$. By imposing
the condition $E\gg Q^{2}/2R\;$ we find that the radius of the ball
must be much smaller than $\sim\varepsilon ^{1/2}/\rho$, where $\rho$
is the charge density. Second,
the size of the system can not be smaller then its one Compton
wavelength (see Eq.~(\ref{defr})). Thus, $R$ is confined to the interval
$$
\gamma\frac{\hbar c}{E}\leq R\ll \sim\frac{\varepsilon ^{1/2}}{\rho}.
$$        

In order to minimize the right hand side of Eq.~(\ref{extb}) we
replace $R$ by its minimal value $R_{min}=\gamma\hbar/E=(3\gamma\hbar
c/4\pi\varepsilon)^{1/4}$. This implies a new bound on the entropy
density: 
\begin{equation}
s(\varepsilon _{in},...)\leq a\left(\frac{\varepsilon _{in}}
{\hbar c}\right)^{3/4},
\label{general}
\end{equation}
where $a=(12\gamma\pi ^{3})^{1/4}$ is a dimensionless
constant. Multiplying both sides by 
the volume of the system we obtain a new extensive entropy bound for
charged systems:
\begin{equation}
S\leq \frac{2\pi\gamma ^{1/4}}{(\hbar
  c)^{3/4}}\left((ER)^{3/4}-\frac{9}{20}\frac{Q^{2}}{(ER)^{1/4}}\right)\;+\;O(Q^{4}). 
\label{exch}
\end{equation}
Note that in the limit $ER\;\rightarrow\;\gamma\hbar c$, our extensive
bound is a bit more liberal than the original bound~(\ref{ebc}). Of course, for $ER\gg
\gamma\hbar c$ the bound~(\ref{exch}) is much tighter. 

\section{An extensive entropy bound for rotating systems}

Using arguments similar to those that motivated Bekenstein to propose the
universal upper entropy bound~(\ref{bb}), Hod~\cite{Hod1} was able to
infer a tighter entropy bound for rotating systems:
\begin{equation}
S\leq\frac{2\pi ER}{\hbar c}\left(1-\frac{J^{2}}{E^{2}R^{2}}\right)^{1/2}\;,
\label{rotbo}
\end{equation}
where $J$ is the total angular momentum (spin) of the
system. Similarly to the arguments following Eq.~(\ref{ebc}), one can
derive this bound from dimensional considerations assuming the
spherical entropy bound~(\ref{sb}). 
Let us now consider the applications of this bound to extensive
systems.

It is well known that in a state of thermal equilibrium, only a
uniform rotation of a body as a whole is possible~\cite{LanLif}. The
total energy of a rotating (in our case spherical) body may be written
as the sum of 
its internal energy, $E_{in}$, and its kinetic energy of rotation:
$E=E_{in}+J^{2}/2I$, where 
$I$ is the moment of inertia of a spherical ball. In general, rotation
changes the distribution of mass in the system, and therefore both $I$
and $E_{in}$ are functions of $J$. However, in this section, we will
consider only a 
sufficiently slow rotation ($E\gg J^{2}/2I$), so that $I$ and $E_{in}$
may be regarded as constants independent of $J$. In this case,
Eq.~(\ref{rotbo}) implies (see~\cite{BekMayo})
\begin{equation}
S\leq \frac{2\pi R}{\hbar c}\left(E-\frac{J^{2}}{2\mu
R^{2}}\right)+O(J^{4})\;, 
\label{aaa}
\end{equation}
where $E$ has been replaced with the rest mass, $\mu$, in the
denominator. 

Now, since the entropy is not a function of the total energy $E$, but
a function of the internal energy, $E_{in}$, both the Bekenstein
bound~(\ref{bb}), as well as Eq.~(\ref{aaa}), imply
Eq.~(\ref{nonextb}), where in this section $E_{in}$ is the internal
energy of a (spherical) rotating body. Thus, the bound given in
Eq.~(\ref{general}) 
is valid also for rotating systems, assuming the entropy is an
extensive function of $E_{in}$. 

For a spherical ball with rest mass,
$\mu$, the moment of inertia has the value, ${2\over 5}\mu R^{2}$, so
that the internal energy for given $E$ and $R$ is:
$E_{in}=E-2J^{2}/5\mu R^{2}$. Substituting the value of the internal
energy density $\varepsilon _{in}=E_{in}/V$ in Eq.~(\ref{general}),
and then multiplying both sides by the volume, $V$, we obtain a new
extensive bound for rotating systems:
\begin{equation}
S\leq\frac{2\pi\gamma ^{1/4}}{(\hbar c)^{3/4}}
\left[(ER)^{3/4}-\frac{3}{10}\frac{J^{2}}{(E R)^{5/4}}\right]
+O(J^{4})\;,
\label{ggg}
\end{equation}
where we have replaced $\mu\;\rightarrow\;E$ in the denominator. Again,
in the limit $ER\;\rightarrow\;\gamma\hbar c$, the extensive
bound is a bit more liberal than the original bound~(\ref{aaa}).   

\section{Applications of the SEB on extensive charged-rotating systems} 

Another interesting question that we would like to address here
concerns the application of the SEB on extensive
systems (not necessarily in flat spacetime). In our previous work it
was shown that the Holographic principle predicts the CEB for extensive 
systems. However, in our derivation we did not
include the net charge (or angular momentum) of the system. 
As it will be shown in this section (without pretending to any
rigor), for charged rotating systems the causal bound takes a
different form. 

Applying the SEB~(\ref{sb}) on extensive systems
leads to the following bound on the entropy density of the system:
\begin{equation}
s(\varepsilon _{in},...)\leq\frac{1}{R},
\label{extext}
\end{equation}
where in this section we set $c=G=\hbar=1$ and we will stress only
functional dependence, while ignoring numerical factors.
Now, since $\varepsilon _{in}$ and $R$ can be considered to be
independent, one can fix $\varepsilon _{in}$ and increase $R$.
However, by increasing $R$ and at the same time keeping $\varepsilon _{in}$
constant, one reduces the ratio $R/E$. Therefore,
the maximum possible value of $R$ occurs when the system becomes a 
(charged-rotating) black hole.

For the Kerr-Newman black hole, the horizon is located at 
\begin{equation}
r_{+}=E+(E^{2}-Q^{2}-J^{2}/E^{2})^{1/2},
\end{equation}       
where here $E$ represents the black hole's mass. The horizon area of
such a black hole is: $A=4\pi(r^{2}_{+}+J^{2}/E^{2})$. Thus, we conclude
that $R \geq\sqrt{A/4\pi}$. In terms of $\varepsilon _{in}$  this
condition can be written in the form
\begin{equation}
R < \frac{1}{\varepsilon _{in}^{1/2}}+O(Q^{2},J^{2})\;.
\end{equation}
Next, substituting the maximum value of $R$ in Eq.~(\ref{extext}) we
infer the entropy density bound
\begin{equation}
s(\varepsilon _{in},...)\leq \sqrt{\varepsilon _{in}}+ O(Q^{2},J^{2})\;.
\end{equation}
Now, since $s$ depends on $Q$ and $J$ only through $\varepsilon _{in}$,
we infer 
 \begin{equation}
s(\varepsilon _{in},...)\leq \sqrt{\varepsilon _{in}}\;.
\label{dd}
\end{equation}
Finally, multiply both sides of the equation above by the volume of
the system, we find that for extensive, charged and rotating systems,
the SEB implies:
\begin{equation}
S(E,V,Q,J,...) < \sqrt{EV}
\left[1-\frac{3Q^{2}}{10ER}-\frac{J^{2}}{5E^{2}R^{2}}\right].
\label{cu}
\end{equation}    
where we have substitute $E_{in}=E-2J^{2}/5\mu R^{2}-3Q^{2}/5R$. This
result will be generalized in the next section for the case
of non-extensive systems.

\section{Discussion: generalization of the Bekenstein and causal
entropy bounds for non-extensive charged and rotating systems}

In this paper we have proposed two new entropy bounds for an isolated
extensive, charged and rotating system. It was assumed that the
entropy of the system is an extensive function of the internal energy. 
Furthermore, our results applicable only for $E\gg Q^{2}/R$ and $E\gg
J^{2}/\mu R^{2}$;
otherwise the particles will be concentrated on the edge of
the system, and thus the entropy will not be an extensive function of
$E_{in}$. However, if the gravitational force is taken into account, it is
possible to imagine extensive systems with relatively high charge and
angular momentum. For such systems our bounds in the previous sections
can be applied.   

Although we have used Eqs.~(\ref{ebc},\ref{rotbo}) to derive the
bounds given in Eqs.~(\ref{exch},\ref{ggg}), it was not really
mandatory. We could obtain the same extensive bounds assuming 
{\it only} the BEB~(\ref{bb}). This is a consequence of
the following ``generalization'' of the Bekenstein bound.

The entropy of charged rotating systems depends on the internal
energy, $E_{in}$~\cite{LanLif}. Thus, the entropy can be written as a
function $S(E_{in},V,...)$. The dots indicates possible dependence on other
parameters such as the number of particles, but {\it not} on the
charge or the angular momentum of the system, because their dependence
is included in $E_{in}$. Now, the total energy of the system, can be
written in the form: $E=E_{in}+\Delta (Q,J,V)$, where $\Delta$ is the
sum of the Coulomb and rotational energies. Hence, the BEB can be
written in the form: 
\begin{equation}
S(E_{in},V,...)\leq\frac{2\pi E_{in}R}{\hbar c}+\frac{2\pi R}{\hbar c}\Delta(Q,J,V).
\end{equation}  
However, the right hand side depends explicitly on $R$, $E_{in}$,$Q$ and $J$, whereas
the left hand side depends only on $E_{in}$ and $R$ (and maybe on some
other parameters that are not relevant for our argument). Thus, we
infer that
\begin{equation}
S\leq \frac{2\pi R E_{in}}{\hbar c}.
\label{genbek}
\end{equation} 
This is a generalized version of the BEB~(\ref{bb}).

The bound~(\ref{genbek}) for lots of systems is tighter than
Eq.~(\ref{ebc}) or Eq.~(\ref{rotbo}). For charged non-rotating
spherical systems it coincides with Eq.~(\ref{ebc}), when the charge is
uniformly spread on the edge of the system (a spherical
shell). Although the bound~(\ref{genbek}) is (usually) smaller,
the bounds in Eq.~(\ref{ebc}) and Eq.~(\ref{rotbo}) have the advantage
that one does not have to know $E_{in}$, but only the total energy,
charge and angular momentum.

The exact same arguments that motivated us to propose
Eq.~(\ref{genbek}), lead us to generalize the CEB in
the following form (we set $k_{B}=c=G=1$ and ignore numerical factors):
\begin{equation}
S<\sqrt{E_{in}V}.
\label{zz}
\end{equation}  
This is reduced to Eq.~(\ref{cu}) when the charge and angular momentum are
relatively small. However, Eq.~(\ref{zz}) can be applied also for
highly charged-rotating systems.

In our previous work, we have obtained the CEB for
extensive systems assuming the the SEB~(\ref{sb}). However, if one
includes the logarithmic correction to the SEB
(see Eq.~(\ref{BH})), than the same arguments that led to
Eq.~(\ref{dd}), implies the following correction term (for simplicity
we assume $J=Q=0$ so that $\varepsilon_{in}=\varepsilon$):
\begin{equation}
s(\varepsilon,...)\leq \sqrt{\varepsilon }
+k\varepsilon ^{3/2}\ln\varepsilon\;. 
\end{equation}  
Thus, Eq.~(\ref{zz}) is replaced by
\begin{equation}
S<\sqrt{EV}+k\frac{E^{3/2}}{V^{1/2}}\ln\left({E\over V}\right).
\label{100}
\end{equation}
The above correction term may be compared with first order correction
to the original definition of the CEB~\cite{BV00,Sasa}. We
leave this comparison for a future work.
   
In conclusion, if the entropy of a black hole includes a logarithmic
correction term (see Eq.~(\ref{BH})), then both the causal and the
Bekenstein entropy bound have to be modified as given in
Eq.~(\ref{100}) and Eq.~(\ref{modbek}), respectively. Furthermore, it
has been shown that one can replace $E\;\rightarrow\;E_{in}$ in the
formulas of these bounds in order to obtain tighter bounds for charged
and rotating systems.

\section{Acknowledgments} 
I would like to thank J.~Bekenstein for his influence and inspiration.
I would also like to thank D.~Page for fruitful discussions and to
V.~Frolov for helpful comments.
The author is also grateful to the Killam Trust for its financial
support.

\end{document}